\documentclass[twocolumn]{article}
\usepackage{times}
\usepackage{epsfig}
\usepackage{graphicx}
\usepackage{amsmath}
\usepackage{amssymb}
\usepackage{booktabs}
\usepackage{xspace}

\usepackage{bbm}
\usepackage[inline]{enumitem}

\usepackage{pdfpages} % include pdfs
\usepackage{pgffor} % for loops

\makeatletter
\AtBeginDocument{\let\LS@rot\@undefined}
\makeatother

\def\supplementfilename{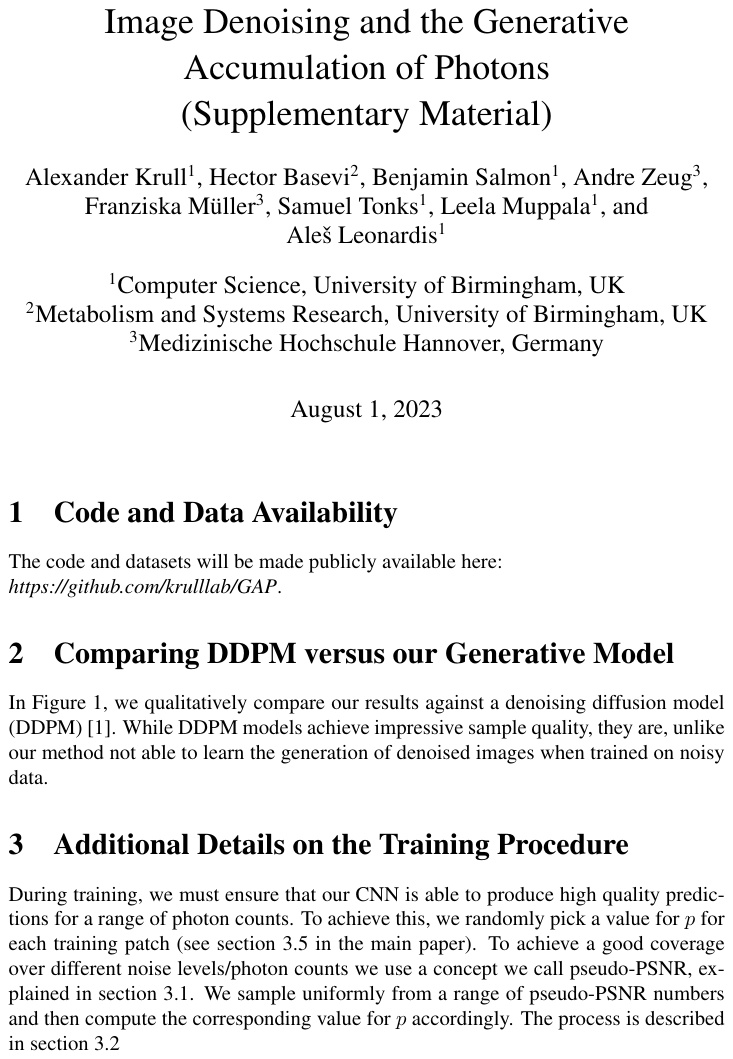}

\pdfximage{\supplementfilename}
\def\numbersupplementpages{\the\pdflastximagepages}

\newif\ifarXiv
\arXivtrue 
\usepackage{authblk}

\def\onedot{.\@\xspace}

\def\eg{\emph{e.g}\onedot} 
\def\ie{\emph{i.e}\onedot}

\def\etal{\emph{et al}\onedot}

\newcommand{\sig}{\mathbf{s}}
\newcommand{\sigpix}{s}

\newcommand{\obs}{\mathbf{x}}
\newcommand{\obspix}{x}

\newcommand{\seq}{\mathbf{i}}

\newcommand{\pars}{\mathbf{\theta}}

\newcommand{\indic}{\mathbbm{1}}

\usepackage{multirow}
\usepackage[normalem]{ulem}
\useunder{\uline}{\ul}{}

\newcommand\figprocess{
\begin{figure*}[h!]
    \centering
    \includegraphics[width=1\linewidth]{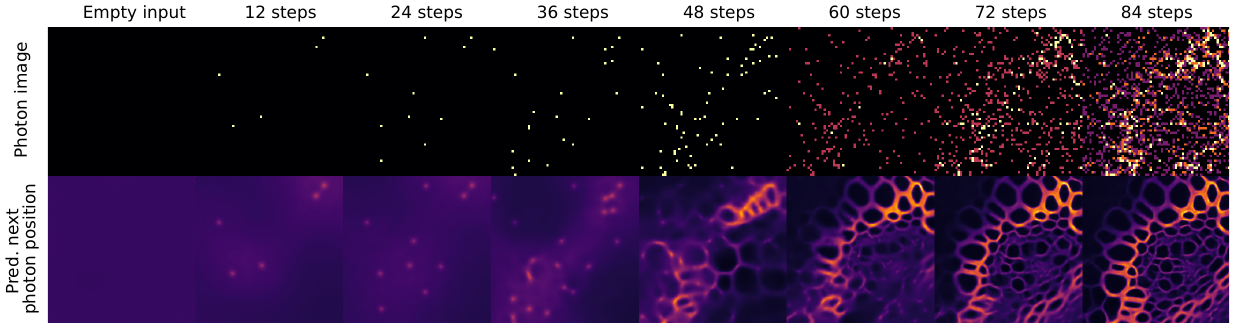}
    \caption{
    \textbf{Generative accumulation of photons (GAP):} 
    Starting with an empty canvas, our method repeatedly predicts a map of probabilities of where the next photon might arrive and uses it to randomly place photons.
    We show the process for the \emph{Neuro-PC} and \emph{Conv-PC} datasets. 
    Photon images have been down-sampled for better visibility.
    }
    \label{fig:gap}
\end{figure*}
}

\newcommand\figexpl{
\begin{figure}[h!]
    \centering
    \includegraphics[width=1\linewidth]{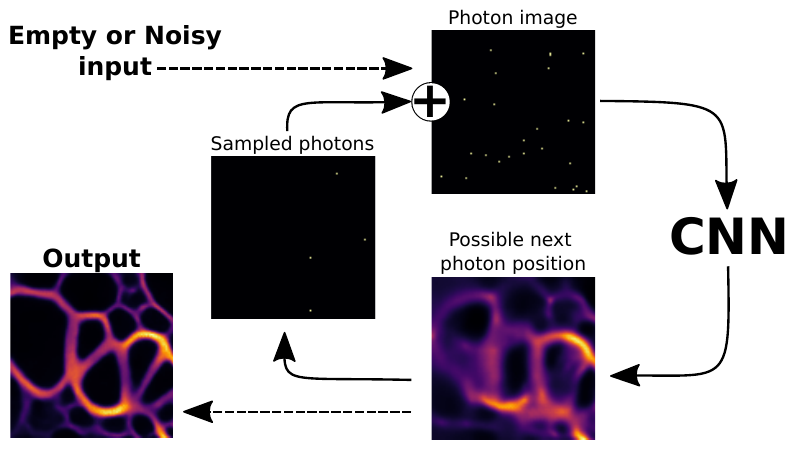}
    \caption{
    \textbf{Sampling algorithm:} 
    Starting with a noisy or empty image, our method repeatedly predicts a map of where the next photon might arrive and uses it to randomly place a small number of photons.
    }
    \label{fig:explainer}
    \vspace{-4mm}
\end{figure}
}

\newcommand\figsplit{
\begin{figure}[h!]
    \centering
    \includegraphics[width=1\linewidth]{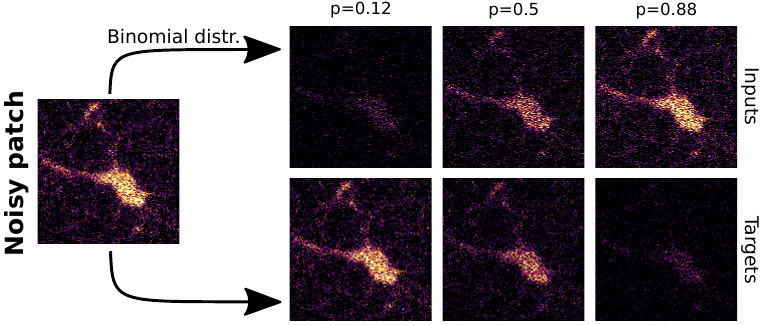}
    \caption{
    \textbf{Photon splitting:}
    We propose a new way of generating training pairs that requires only noisy data.
    We split the noisy image by randomly assigning photons to the input or target image.
    The number of photons assigned to the input is drawn from a binomial distribution, with a  parameter $p$ controlling the noise level. Remaining photons are assigned to the target.
    }
    \label{fig:split}
\end{figure}
}

\newcommand\figsamples{
\begin{figure}[h!]
    \centering
    \includegraphics[width=\linewidth]{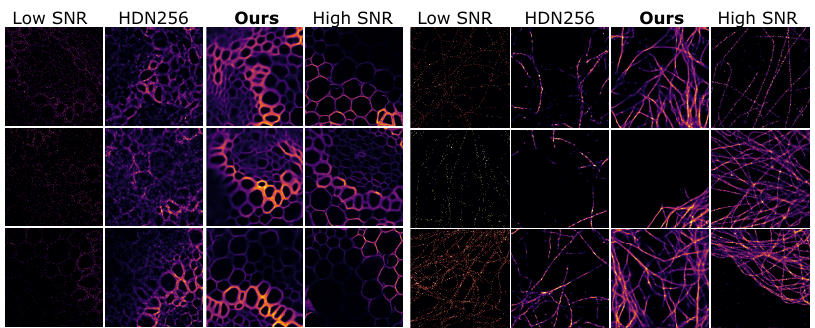}
    \caption{
    \textbf{Comparison of sample quality:}
    We show randomly selected generated samples for the \emph{Conv-PC} and \emph{MT-SM} datasets compared to randomly cropped real high- and low-SNR patches.
    }
    \label{fig:samples}
\end{figure}
}

\newcommand\figdivDenoising{
\begin{figure}[h!]
    \centering
    \includegraphics[width=1\linewidth]{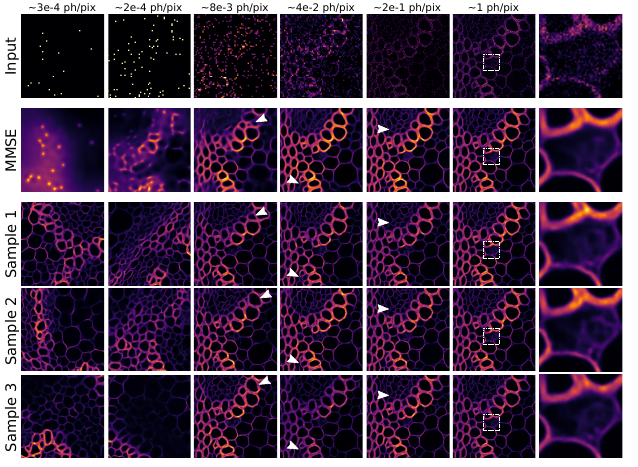}
    \caption{
    \textbf{Diverse solutions for shot noise removal:} 
    GAP models can be used to remove shot noise by taking the noisy image as starting point and sequentially adding additional photons, until a clean image is produced.
    Less noisy inputs lead to less diverse predictions as more information about the clean image becomes available.
    The last column depicts a zoomed in region indicated by the dashed box.
    Arrows highlight structural differences in the samples.
    }
    \label{fig:divdenoising}
    \vspace{-1mm}
\end{figure}
}

\newcommand\figresultsm{
\begin{figure*}[h!]
    \centering
    \includegraphics[width=1\linewidth]{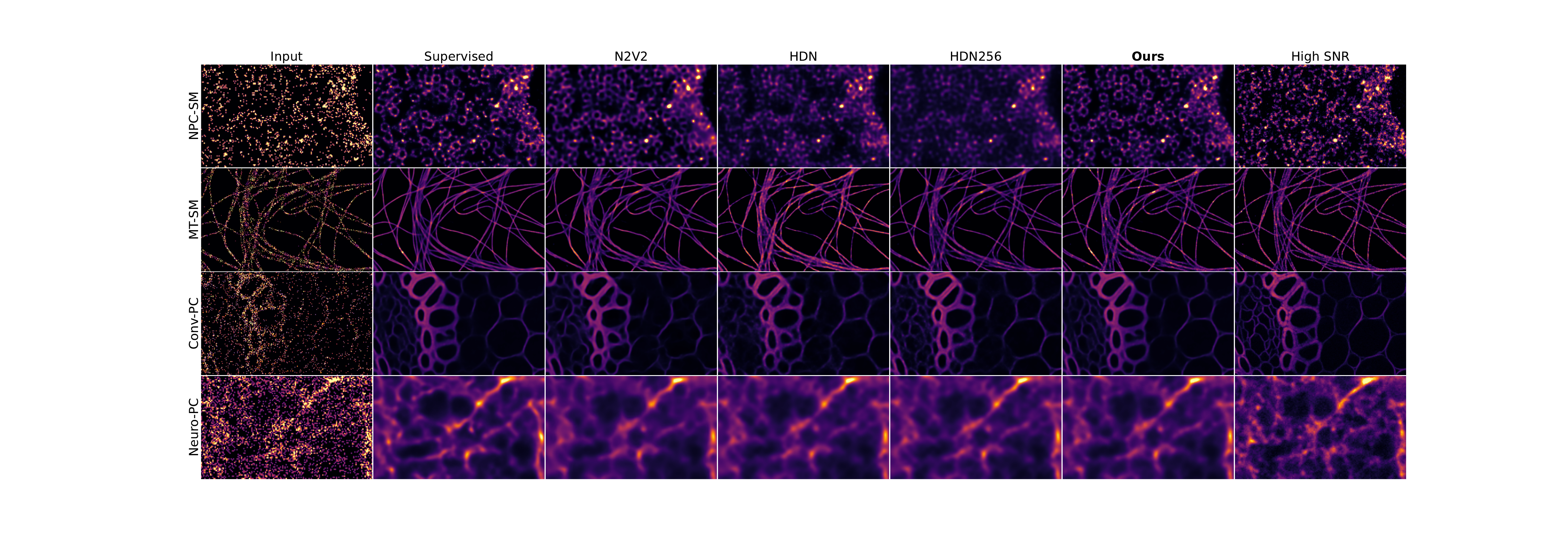}
    \caption{
    \textbf{Qualitative denoising results:} 
    We show MMSE denoising results for all datasets and methods.
    All baselines except for the supervised one have been trained purely on low-SNR input data. 
    }
    \vspace{-4mm}
    \label{fig:resultsm}
\end{figure*}
}

\newcommand\tabpsnr{
\setlength{\tabcolsep}{5pt}
\begin{table}[]
\caption{Average peak signal to noise ratios (PSNRs) in dB (higher is better).
\label{tab:psnr}
}
\begin{tabular}{l|rrrrr}
         & \multicolumn{1}{c}{Superv.} & \multicolumn{1}{c}{N2V} & \multicolumn{1}{c}{HDN} & \multicolumn{1}{c}{HDN256} & \multicolumn{1}{c}{Ours} \\ \hline
NPC-SM   & 39.09                       & 39.00                   & 39.01                   & 38.73                      & \textbf{39.17}           \\
MT-SM    & 36.59                       & 36.17                   & 34.32                   & 36.43                      & \textbf{36.64}           \\
Conv-PC  & 20.42                       & 23.85                   & 23.77                   & 24.32                      & \textbf{24.84}           \\
Neuro-PC & 31.74                       & 32.61                   & 32.41                   & 32.56                      & \textbf{32.63}          
\end{tabular}
\vspace{-1mm}
\end{table}
}

\newcommand\tabfid{
\begin{table}[]
\centering
\caption{We show the FID score~\cite{heusel2017gans} of 10k generated images compared to 10k random crops of high-SNR data.
Note that all methods have been trained on low-SNR training data.
\label{tab:fid}
}
\begin{tabular}{lrrrr}
                              & \multicolumn{2}{c}{64x64 pixels}                               & \multicolumn{2}{c}{256x256 pixels}                    \\
\multicolumn{1}{l|}{}         & \multicolumn{1}{c}{HDN} & \multicolumn{1}{c|}{Ours}            & \multicolumn{1}{c}{HDN256} & \multicolumn{1}{c}{Ours} \\ \hline
\multicolumn{1}{l|}{NPC-SM}   & \textbf{146.48}         & \multicolumn{1}{r|}{146.92}          & 204.34                     & \textbf{95.32}           \\
\multicolumn{1}{l|}{MT-SM}    & \textbf{130.20}         & \multicolumn{1}{r|}{138.72}          & 132.55                     & \textbf{84.49}            \\
\multicolumn{1}{l|}{Conv-PC}  & 200.70                  & \multicolumn{1}{r|}{\textbf{134.20}} & 163.28                     & \textbf{86.82}           \\
\multicolumn{1}{l|}{Neuro-PC} & \textbf{111.23}         & \multicolumn{1}{r|}{115.06}          & 77.90                      & \textbf{64.16}          
\end{tabular}
\vspace{-1mm}
\end{table}
}

\begin{document}

%%%%%%%%% TITLE
\title{Image Denoising and the Generative Accumulation of Photons}

\author[1]{Alexander~Krull}
\author[2]{Hector~Basevi}
\author[1]{Benjamin~Salmon}
\author[3]{Andre~Zeug}
\author[3]{Franziska~Müller}
\author[1]{Samuel~Tonks}
\author[1]{Leela~Muppala}
\author[1]{Ale\v{s}~Leonardis}

\affil[1]{Computer Science, University of Birmingham, UK}
\affil[2]{Metabolism and Systems Research, University of Birmingham, UK}
\affil[3]{Medizinische Hochschule Hannover, Germany}

\maketitle
\thispagestyle{empty}

%%%%%%%%% ABSTRACT
\begin{abstract}
% INTRODUCE GAP
We present a fresh perspective on shot noise corrupted images and noise removal.
By viewing image formation as the sequential accumulation of photons on a detector grid, we show that a network trained to predict where the next photon could arrive is in fact solving the minimum mean square error (MMSE) denoising task.
This new perspective allows us to make three contributions:
\begin{enumerate*}[label=\roman*.]
    \item We present a new strategy for self-supervised denoising,
    \item We present a new method for sampling from the posterior of possible solutions by iteratively sampling and adding small numbers of photons to the image.
    \item We derive a full generative model by starting this process from an empty canvas.
\end{enumerate*}
We call this approach generative accumulation of photons (GAP).
We evaluate our method quantitatively and qualitatively on 4 new fluorescence microscopy datasets, which will be made available to the community.
We find that it outperforms supervised, self-supervised and unsupervised baselines or performs on-par.
\end{abstract}
%%%%%%%%% BODY TEXT
\vspace{-3mm}

% ---------------------------------------------------
\section{Introduction}
% ---------------------------------------------------
\figprocess
Scientific imaging techniques such as fluorescence microscopy have to limit the amount of light used to avoid damaging or destroying their sample~\cite{icha2017phototoxicity}.
As a result, the recorded images inevitably suffer from a certain degree of noise which has to be addressed in the downstream analysis.
Images can be subject to a variety of different types~\cite{laine2021imaging} of noise which can be alleviated by various technical means (\eg \cite{denvir2003electron}).
However, there is a type of noise which is physically inevitable for most imaging setups in low-light conditions.
It is referred to as Poisson \emph{shot noise}.

Shot noise is the result of the particle nature of light.
Even high-end scientific detectors and cameras that can accurately count the precise number of photons hitting each pixel cannot record a noise-free image.
For a given light intensity the number of photons arriving at the detector is itself inherently random and follows a Poisson distribution.
The effect is especially severe in microscopy applications, operating in low-light conditions.

The last decade has seen a number of deep learning-based computational methods designed to reduce noise after images have been recorded in order to allow for improved analysis of the data~\cite{laine2021imaging}.
One of the first proposed methods, known as content-aware image restoration~(CARE)~\cite{weigert2018content}, is based on training convolutional neural networks (CNNs) to learn a mapping from noisy images to clean images.
Unfortunately, the method requires pairs of corresponding noisy and clean images during training, which can be hard to acquire in practice, rendering it inapplicable in many situations. 
However, other works have expanded on this line of research, enabling training with noisy image pairs~\cite{lehtinen2018noise2noise} and even with unpaired noisy images, \eg~\cite{krull2019noise2void,batson2019noise2self,Broaddus2020_ISBI, quan2020self2self}.

While achieving impressive results, these supervised and self-supervised methods share a common shortcoming:
Denoising is inherently an ill-posed problem and given an image corrupted by a substantial amount of noise, it is not generally possible to recover the true underlying clean image.
In fact, there is a posterior distribution of possible solutions that all might have led to the original noisy observation.
When we view denoising as a regression problem like~\cite{weigert2018content} aiming to learn a mapping to the clean image, we are in fact learning a mapping to a compromise between possible solutions, which may itself look different (often being more blurry) from real clean images.

This problem has been explored by Prakash~\etal~\cite{prakash2020fully,prakash2022interpretable}, who proposed the idea of \emph{diversity denoising} based on a variational autoencoder~\cite{kingma2014auto} (VAE).
Instead of producing a single solution for each noisy image, Prakash\etal are able to sample possible solutions from an approximate posterior distribution of clean images.

Here, we take an entirely new perspective, focusing on shot noise and the denoising of shot noise corrupted images.
Instead of viewing shot noise as a secondary corruption process applied to a clean image, we understand image formation as the sequential accumulation of photons and see any measured shot noise-affected image as a result of this process.
We call this approach generative accumulation of photons (GAP).

We train a CNN to take a shot noise-affected image as input and predict a probability distribution over where the next photon might arrive.
We show that, for normalized images, predicting the next photon position is identical to denoising the image.

Based on this insight, we derive a novel method of training a self-supervised model for image denoising.
Additionally, by understanding image generation as the accumulation of photons, we describe a new method for diversity denoising: we iteratively predict the distribution of the next photon position and randomly sample photons accordingly.
Finally, by starting with an empty photon-free canvas, we are able to derive a full generative image model.
The process is illustrated in Figures~\ref{fig:gap}~and~\ref{fig:explainer}.

We introduce four new shot noise-corrupted microscopy datasets for evaluation that will be made available to the community.
We evaluate our method quantitatively and qualitatively and find that it yields competitive results.

% ---------------------------------------------------
\section{Related Work}
% ---------------------------------------------------
\subsection{Supervised denoising}
Supervised denoising methods usually train CNNs using pairs of noisy and clean images to learn a mapping between the two. 
A common choice for the loss function is the mean square error (MSE) between the prediction and clean ground truth.
Considering that there is a distribution of possible solutions, minimising the MSE loss corresponds to finding the expected value.
Unfortunately obtaining clean ground truth data can be challenging or impossible for many applications and so supervised methods are often not applicable in the context of scientific imaging.

Lehtinen \etal introduced \emph{Noise2Noise}~\cite{lehtinen2018noise2noise}, a partial solution to this problem. 
They showed that it is possible to replace the clean ground truth target with a second noisy version which might be more readily available.
A network trained with this type of data will still find the same MMSE solution.
While this presented a big step forward with respect to applicability, Noise2Noise still requires training pairs, which have to be collected for this purpose.

\subsection{Self-supervised blind-spot denoising}
Self-supervised blind-spot methods~\cite{krull2019noise2void, batson2019noise2self,hock2022n2v2} suggest a training strategy that can do without paired training data, \ie allowing training directly on the data that should be denoised, while still obtaining the same MMSE solution.
The main idea is to block out individual pixels in order to use them as noisy targets (similar to Noise2Noise).  
These strategies rely on the assumption that imaging noise is conditionally pixel-independent given the underlying clean signal, making it not possible to predict the noise in a pixel from its surroundings.
The downside of this approach is that, when making a prediction for a pixel, the network cannot make use of the pixel value itself, thus it is not making optimal use of the available information.

Our photon-based self-supervised denoising strategy is related to the blind-spot idea in that it removes part of the input image to use it as the target.
However, instead of removing pixels, we are only removing individual photons, which means we are not facing the same problem of disregarded information.

\subsection{VAE-based denoising}
Another approach to image denoising has been suggested in~\cite{prakash2020fully}.
The core idea is to use a variational autoencoder to describe the distribution of noisy images.
By including a statistical model of the imaging noise as part of the decoder, the method allows us to:
\begin{enumerate*}[label=\roman*.]
\item sample from an approximate posterior distribution of possible clean images, and
\item to sample clean images from scratch, functioning as a full generative model.
\end{enumerate*}

An extended method with a more powerful network architecture was presented in~\cite{prakash2022interpretable} under the name HDN.
We see this method as our main competitor as it can be trained from unpaired noisy data and, similarly to our method, can function as a generative model.

Unlike GAP which produces an MMSE denoising result in a single step, HDN produces MMSE results by repeated sampling and averaging from the posterior distribution.

\subsection{Generative Image Models}
Generative image models aim to describe a probability distribution over images, a highly challenging task, due to:
\begin{enumerate*}[label=\roman*.]
\item the high dimensionality of the random variable (the number of pixels) and
\item \label{enum_correlations} due to the complex higher-order correlations between pixel values at different locations.
\end{enumerate*}
As a result of \ref{enum_correlations} the distribution cannot easily be factorised into lower order terms and attempts to factorise using methods such as Markov random fields (MRFs)~\cite{li2009markov} have led to overly simplistic results that do not realistically describe the image distribution. 

In recent years, a number of approaches to this problem have been highly successful.
Latent variable models, such as generative adversarial networks (GANs)~\cite{goodfellow2014generative} and VAEs~\cite{kingma2014auto}, or normalising~flows~\cite{rezende2015variational}, describe difficult distributions indirectly by starting with an easily modelled high dimensional latent variable (usually following a normal distribution) which is then deformed using convolutional neural networks (CNNs) to describe the distributions of interest.

A different approach to this is autoregressive modelling, as proposed by Van~Oord~\etal~\cite{van2016pixel}.
By viewing image generation as a sequential process in which the pixels of an image are thought to be generated one-at-a-time conditioned on all previous pixels.
In this setup, the whole model can be formulated as a product of 1D conditional distributions over each pixel's intensity value.
Our method can be viewed as an autoregressive approach as we model image generation as a sequential process. 
However, we sample images by sequentially placing individual photons instead of drawing pixel values.

Finally, the current state-of-the-art approach to image modelling, {\em denoising diffusion models}~\cite{ho2020denoising}, follows a similar approach by describing image generation as a sequence of steps.
The process is inspired by physics and considers an image as a particle in a high dimensional space, diffusing away from its original position according to some noise distribution.
To generate an image the denoising diffusion approach reverses the diffusion process by applying a sequence of denoising steps.

Denoising diffusion models iteratively reverse a diffusion process on clean images which typically involves Gaussian noise. This noise can be applied directly to the image \cite{dickstein2015deep,ho2020denoising}, or instead to a latent representation of the image corresponding to a pre-existing autoencoder \cite{rombach2022high}. In both cases, the diffusion noise distribution is unrelated to the noise distribution of the training data. The diffusion model learns to sample from the \emph{noisy} training data distribution and so its samples contain this noise. In contrast, GAP learns to sample from a noisy training data distribution \emph{and} to denoise this distribution. In addition, every iteration of GAP results in a physically valid noisy image.

Recent works have explored generalisations of diffusion models to broader families of corruption processes. Bansal \etal \cite{bansal2022cold} focused on deterministic image corruptions. Daras \etal \cite{daras2022soft} focused on image corruptions which are linear with respect to the clean image. GAP focuses on shot noise, which is neither deterministic nor linear.

% ---------------------------------------------------
\section{Method} 
\label{sec:methods}
% ---------------------------------------------------
\figexpl
% ...................................................
\subsection{Image Formation and Shot Noise}
% ...................................................
When we record an image, we usually project light onto a digital sensor, such as a CMOS or a CCD chip.\footnote{
Some imaging technologies, especially those capable of counting photons, work by scanning the sample and recording one pixel at a time. Since this does not affect our model we will focus our explanation on camera-based systems for simplicity.}
These chips contain many detector elements measuring the amount of light arriving at different locations on the chip.
In our simplified model we assume that each of these detector elements corresponds to one pixel of the final image.
When measuring the amount of light in each pixel, we treat light as discrete particles called photons.
In an ideal case with a perfect detector, each pixel value in the final image corresponds to the number of photons that fell onto the pixel.

The result of this process is a \emph{shot noise} corrupted image $\obs = (\obspix_1, \dots, \obspix_n)$, where the photon count $\obspix_i$ in each pixel $i$, is independently drawn from a Poisson distribution
\begin{equation}
    p(\obspix_i|\sigpix_i)= 
    \frac
    {\sigpix_i^{\obspix_i} \exp({-\sigpix_i})}
    {\obspix_i !}
    , 
\end{equation}
where $\sigpix_i$ refers to the expected number of photons hitting the pixel $i$ during the exposure, \ie to the light intensity at the pixel -- the quantity we were originally interested in measuring.
We will refer to the vector $\sig=(\sigpix_1, \dots, \sigpix_n)$ as the signal or as a clean image.

Since photons are hitting each pixel independently given a signal, we can describe the probability of observing a noisy image $\obs$ given a signal $\sig$ as 
\begin{equation}
    p(\obs|\sig) = 
    \prod_{i=1}^{n}
    p(\obspix_i|\sigpix_i)
    .
    \label{eq:shot_noise}
\end{equation}
We can now think of image formation as a two-step process.
We can imagine an image being created by first drawing a clean image $\sig$ from a distribution $p(\sig)$ and then applying shot noise by drawing photon counts from Eq.~\ref{eq:shot_noise} to create the shot noise-corrupted version.

% ...................................................
\subsection{The Denoising Task}
% ...................................................
Given noisy observation $\obs$, denoising is defined as finding an estimate $\hat \sig$ for the unknown clean image $\sig$. 

However, considering the process of image generation described above, finding the true signal may not be possible since many clean images can lead to the same noisy observation.
We can use Bayes' theorem to write down a posterior distribution over possible clean images for a given noisy observation
\begin{equation}
    p(\sig|\obs) \propto p(\obs|\sig) p(\sig)
    .
    \label{eq:posterior}
\end{equation}

Deep learning-based approaches (\eg \cite{weigert2018content, krull2019noise2void}) often view denoising as a regression problem and use CNNs to try to directly learn a mapping from $\obs$ to $\sig$.
When such methods are trained with a mean squared error (MSE) loss function the optimal solution is the expectation
\begin{equation}
    \hat{\sig} = \int p(\sig|\obs) \sig \: d \sig
    .
    \label{eq:mmse}
\end{equation}
We call this the minimum mean squared error (MMSE) solution.
This is a sensible way to find an estimate, but we should be aware that it constitutes a compromise between all possible $\sig$.
% Depending on the amount of noise that is present in the image, the MMSE estimate can be very blurry and lacking detail as it is the compromise between a wide range of possible true clean images.
\figsplit

% ...................................................
\subsection{Image Generation from Photon Sequences}
% ...................................................
Here, we take an alternative view on image generation.
Instead of thinking of our pixel values as being drawn from a Poisson distribution, we will derive an equivalent description, viewing image generation as a sequential process.
Remembering that our observation $\obs$ is created by photons hitting our detector, we can imagine that it was created  by an ordered sequence of photons $\seq = (i_1, \dots, i_T)$, where $i_t$ is the index of the pixel where photon $t$ hit the detector.
The index $t$ simply refers to the position of the photon in the sequence, with the first photon arriving at $t=1$ and the last one at $t=T$.

Assuming a known sequence of photons, we compute the resulting image by counting the number of photons hitting each pixel as
\begin{equation}
    \obspix_i = \sum_{t=1}^T \indic(i_t=i),
    \label{eq:counting}
\end{equation}
where $\indic(i_t=i)$ is the indicator function.

Considering a given signal $\sig$ and a known number $T$ of photons, we can compute the probability of a sequence $\seq$ as
\begin{equation}
p(\seq|\sig,T) = 
\begin{cases} 
      \prod_{t=1}^T p(i=i_t|\sig) 
    & T = |\seq|  \\
    0 & T \neq |\seq| \\
\end{cases},
\end{equation}
with the probability being 0 where the length $|\seq|$ of the sequence does not match the number of photons $T$. 
Since the photons hit the detector independently, their position in the sequence does not matter and we can rewrite the probability as a product over pixels as 
\begin{equation}
p(\seq|\sig,T) = 
\begin{cases} 
      \prod_{i=1}^n p(i|\sig)^{\obspix_i}
    & T = |\seq|  \\
    0 & T \neq |\seq| \\
\end{cases}
\label{eq:known_sig}
.
\end{equation}

where the probability $p(i|\sig)$ for a photon to hit a particular pixel $i$, given the signal, should be proportional to the light intensity at the pixel.
Thus, we can compute it as the normalised signal at that pixel
\begin{equation}
    p(i_t=i|\sig) = 
    \frac
    {\sigpix_i}
    {\sum_{j=1}^n \sigpix_j}
    .
    \label{eq:nomalised_signal}
\end{equation}

However, if the clean signal is unknown the distribution will no longer factorise as easily as Eq.~\ref{eq:known_sig}.
Instead, we have to compute the probability of a sequence as
\begin{equation}
    p(\seq|T) = 
    \begin{cases} 
      \prod_{t=1}^T p(i=i_t|i_1, \dots, i_{t-1}, T) & T = |\seq|  \\
    0 & T \neq |\seq| \\
\end{cases}
    ,
    \label{eq:seq}
\end{equation}
where the distribution $p(i=i_t|i_1, \dots, i_{t-1}, T)$ of the next possible photon location now depends on all previous photons.
The order in which photons $i_1, \dots, i_{t-1}$ arrived does not provide any information regarding the next photon position.
By additionally considering that the next photon position does not depend on total photon number $T$ nor on the order of previous photons, we can write
\begin{equation}
    p(i=i_t|i_1, \dots, i_{t-1},T) =
    p(i=i_t|\obs_{t-1})
    ,
    \label{eq:photon_dist}
\end{equation}
where $\obs_{t-1}$ is the observed image at step $t-1$ according to Eq.~\ref{eq:counting}.

Equation~\ref{eq:photon_dist} refers to the distribution over the next possible photon locations given a photon image $\obs_{t-1}$.
Before taking a closer look at how it can be computed, we would like to point out its significance.
Together, Eq.~\ref{eq:seq} and Eq.~\ref{eq:photon_dist} provide not only a way to calculate the probability of a sequence but also an iterative way to sample a sequence of photons and therefore images $\obs_T$.
Furthermore, for large $T$, we can expect $\obs_T$ to approach the clean image $\sig$, when scaled correctly, so that Eq.~\ref{eq:seq} and Eq.~\ref{eq:photon_dist} hold the key to the generation of clean images as well. 

% ...................................................
\subsection{Predicting the Next Photon Location is MMSE Denoising for Normalised Signals}
\label{sec:pred_pl}
% ...................................................
Let us now take a closer look at the distribution of possible next photon locations $p(i=i_t|\obs_{t-1})$.
We can rewrite Eq.~\ref{eq:photon_dist} by marginalising over the unknown signal and using Eq.~\ref{eq:nomalised_signal} as
\begin{align}
    p(i=i_t|\obs_{t-1})
    &= \int p(\sig|\obs_{t-1}) p(i_t=i|\sig, \obs_{t-1}) \: d \sig \\
    &= \int p(\sig|\obs_{t-1}) 
    \frac
    {\sigpix_i}
    {\sum_{j=1}^n \sigpix_j}
    \: d \sig
    .
    \label{eq:marginal}
\end{align}

We can see that the result is a weighted average of the possible normalized signals.
We should expect that the distribution will be high entropy for small $t$, \ie, when we have not yet observed many photons, and that it should become more concentrated and low entropy for large $t$.
For very large $t$, the distribution should approach a normalised version of the signal (Eq.~\ref{eq:nomalised_signal}), because $\obs_t$ will give us more and more information on the underlying signal.

Interestingly, Eq.~\ref{eq:marginal} closely resembles Eq.~\ref{eq:mmse}.
In fact, if we were to consider only normalized signals with $\sum_{j=1}^n \sigpix_j = 1$ the two equations are identical, meaning that the task of predicting the next photon location is identical to denoising the image in an MMSE sense.

We will use a CNN to approximate $f_\pars(\obs_{t-1}) \approx p(i=i_t|\obs_{t-1})$, where $\pars$ are the network parameters.
In section~\ref{sec:learning}, we will discuss how we can train the CNN to achieve this task.

% ...................................................
\subsection{Learning to Predict the Next Photon Location}
\label{sec:learning}
% ...................................................
Based on the insight from section~\ref{sec:pred_pl}, we know that any model trained for MMSE denoising can approximate the distribution over the next photon location $p(i=i_t|\obs_{t-1})$.
Starting with normalised clean training images $\sig^k$, the traditional way of creating training pairs is to simulate the corresponding noisy version $\obs^k$.
We can then train a denoiser network using a standard quadratic loss function, with $\obs^k$ as input and $\sig^k$ as target.

However, in many cases clean data is unavailable. 
Considering the task of predicting the next photon location suggests an alternative self-supervised approach by viewing the problem as a classification task learning the categorical distribution of possible photon positions.
By using a softmax layer over pixels at the output of our network to ensure that outputs sum to one, we can use the standard cross-entropy loss.
In principle, this would require only individual photon positions as target for each training image, just as classifiers are frequently trained using individual class labels for each training example.
We could easily create such training pairs from unpaired noisy images $\obs^k$ by randomly removing a single photon and using it as target.
The corresponding cross entropy loss is
\begin{equation}
    L(\pars) = -
    \sum_{k=1}^m 
    \sum_{i=1}^n
    \ln f_i(\obs^k_{\tiny \mbox{inp}};\pars)
    \obspix^k_{{\tiny \mbox{tar}},i}
    ,
    \label{eq:loss}
\end{equation}
where $m$ is the number of training images, $\obs^k_{\tiny \mbox{inp}}$ is the training image with one photon randomly removed and $\obs^k_{\tiny \mbox{tar}}$ is a one-hot representation of the removed photon position.

However, we require training data at multiple noise levels to enable our network to predict an accurate approximation of $p(i=i_t|\obs_{t-1})$ at different times $t$.
To achieve this, we use a control parameter $p$ and split the image $\obs^k$ into two parts, $\obs^k_{\tiny \mbox{inp}}$ and $\obs^k_{\tiny \mbox{tar}}$.
We can think of this process as simulating a shorter exposure time during image acquisition.
Considering that $\obs^k$ was recorded with a certain exposure time $\tau$, we can imagine what would be the result if we had instead recorded two images consecutively, with the first image being exposed for $p\tau$ and the second being exposed for $(1-p)\tau$.
Considering, that the underlying signal remained fixed during the entire time, each of the photons that make up $\obs^k$ would end up in the first image with probability $p$ and in the second image with probability $(1-p)$.
To efficiently sample a split for parameter value $0<p<1$, we can determine each pixel value $\obspix^k_{{\tiny \mbox{inp}},i}$ by drawing from binomial distribution using $p$ and $\obspix^k_i$ as the distributions parameters, for success probability and number of trials, respectively.
We can then compute the number photons in the target image as $\obspix^k_{{\tiny \mbox{inp}},i} =  \obspix^k_i - \obspix^k_{{\tiny \mbox{inp}},i}$.
By changing the value $p$ we can control the number of photons that are on average assigned to the input or target image respectively. 
The process is illustrated in Figure~\ref{fig:split}.
We use a randomly selected $p$ for each training patch to cover all levels of noise.
We show in the Supplementary material that the loss formulation in Eq.~\ref{eq:loss} can still be used to maximise the likelihood of the training data even when $\obs^k_{\tiny \mbox{tar}}$ is not a one-hot encoding of a single photon position but an image that contains an arbitrary number photons. 
In practice, we use a normalized variant that still maximizes likelihood of the data
\begin{equation}
    L(\pars) =
    -
    \sum_{k=1}^m
    \frac{1}
    {
    {n |\obs^k_{{\tiny \mbox{tar}}}}|}
    \sum_{i=1}^n
    \ln f_i(\obs^k_{\tiny \mbox{inp}};\pars)
    \obspix^k_{{\tiny \mbox{tar}},i}
    ,
    \label{eq:loss_norm}
\end{equation}
where $|\obs^k_{{\tiny \mbox{tar}}}|$ is the sum of photons in $\obs^k_{{\tiny \mbox{tar}}}$.

% ...................................................
\subsection{Inference}
% ...................................................
\noindent {\bf MMSE denoising:}
To compute the MMSE denoising result $\hat{\sig}$ for a noisy input image $\obs$, we can simply apply our trained CNN.
As shown in section~\ref{sec:pred_pl}, the resulting probability distribution corresponds to the MMSE estimate.
However, since our network uses a final softmax layer, the output can only tell us about the normalized pixel intensities and not about the absolute ones.
To obtain a scaled version, that is comparable to the results from N2V2 or HDN, we multiply our output with the number of photons in the input image.

\noindent {\bf Diversity denoising:}
To obtain a sample from the posterior of clean images, given a shot noise corrupted input, we use the iterative procedure illustrated in Figure~\ref{fig:explainer}.
Starting with the original image, we repeatedly apply our network to obtain the distribution for the next photon position and add photons drawn from this distribution.
Even though Eq.~\ref{eq:seq} contains a product over individual photons and we should in principle draw only a single photon at a time, we find that we can add multiple photons simultaneously while maintaining acceptable quality.
In practice, we add 10\% of the current photon count in each step, increase the total number of photons exponentially.
A more detailed description of photon sampling can be found in the Supplementary materials.

\noindent {\bf Image generation:}
To obtain samples from our generative model, we follow the same process as diversity denoising, but start with a blank image.

% ---------------------------------------------------
\section{Experiments}
% ---------------------------------------------------

% ...................................................
\subsection{Network Architecture and Training}
% ...................................................
Here, we will only give a brief overview of the architecture and training procedure used for the experiments on microscopy data.
A more detailed description can be found in the Supplementary material.

For all our experiments on microscopy datasets, we use a modified UNet~\cite{ronneberger2015u} consisting of 6 levels, with a residual block at each level and skip connections.
We use 28 feature channels in the first level and double the number of feature channels at each subsequent level.
All our networks are trained using the \emph{ADAM}~\cite{kingma2015adam} optimizer for 100 epochs.
We use randomly cropped patches of 256$\times$256 pixels, which are augmented 8-fold, using random flips and transpose operations.
We use a batch size of 32.

% ...................................................
\subsection{Baselines}
% ...................................................
\noindent {\bf \emph{Supervised denoising}} uses the same network architecture as our method except for the softmax layer at the end.
It is trained with the same hyperparameters but uses a MSE loss function.
As for our method, we use 8-fold data augmentation.

\noindent {\bf \emph{N2V2}} uses the implementation from~\etal~\cite{hock2022n2v2}, with default hyper parameters and the default 64$\times$64 training patch size.

\noindent {\bf \emph{HDN}} uses the implementation from Prakash \etal~\cite{prakash2022interpretable},
with default hyper parameters and the default 64$\times$64 training patch size.
HDN requires a model of the imaging noise, which is usually trained from data.
Instead, because we know our data contains pure shot noise, we added an analytical Poisson noise model, accounting for shot noise.

\noindent {\bf \emph{HDN256}} uses the implementation from Prakash \etal~\cite{prakash2022interpretable} but with increased network complexity to allow for a fairer comparison to our method.
Specifically, we increase the dimensionality of the latent variables from 32 feature channels to 70, and the number of deterministic filters in the hidden units from 64 to 140.
The method uses 256$\times$256 pixel training patches.
We use the same noise model as for HDN.
\figresultsm

% ---------------------------------------------------
\subsection{Photon Counting Datasets}
% ---------------------------------------------------
While a number of denoising datasets are available in the microscopy domain (\eg~\cite{zhang2019poisson,hagen2021fluorescence}), none of them show purely shot noise corrupted data.
To address this gap, we introduce four new quantitative datasets, including High-SNR ground truth data and one additional qualitative dataset that does not contain ground truth.

We use two photon-counting datasets that will be made available to the community.
As a result, the recorded pixel intensities give a very accurate approximation of the photons hitting each pixel during the exposure.

\noindent {\bf The \emph{Conv-PC} dataset}
We image 5 fields of view (FOV) repeatedly, 512 times at a resolution of 512 $\times$ 512 pixels.
Each of the individual frames contains a substantial amount of shot noise. 
By summing the 512 images for each FOV, we obtain the high-SNR version.
Four FOVs were used as training data for supervised denoising, the remaining one was used as test data.

\noindent {\bf The \emph{Neuro-PC} dataset} contains images of mouse neurons.
The dataset is created from a z-stack of $2048\times2048$ pixels by using $2\times2$ binning in x- and y-direction direction and 4 times binning in the z-direction.
We divided the images into non-overlapping 320$\times$320 regions and rejected empty ones.
To produce the corresponding low-SNR versions we reduced the photon count in each pixel to simulate a 1000-fold shorter exposure by using a binomial distribution with $p=0.001$.
We use every fourth frame as test set and keep the rest as training set for the supervised baseline.
All in all, this amounts to 133 images of size 320$\times$320, 33 of which are test images.

% ---------------------------------------------------
\subsection{Single Molecule Localisation Microscopy}
% ---------------------------------------------------
Single molecule localisation microscopy (SMLM)~\cite{betzig2006imaging} data is produced differently from photon counting data but is subject to the same type of shot noise corruption. It uses a large set of images of the same field of view to detect and localise individual fluorescent emitters in each image.
The resulting emitter locations are then stored in a list and can be binned in x and y to produce a 2D histogram/image containing the number of emitters in each bin/pixel.

\noindent {\bf The \emph{NPC-SM} dataset} was derived from single molecule localisation data published by  Löschberger~\etal~in~\cite{loschberger2012super}.
It shows the arrangement of the \emph{gp210} protein around the nuclear pore complex (NPC).
To create the dataset, we binned the detected emitter locations using a bin size of 20nm $\times$ 20nm to produce the high-SNR data. 
To produce the corresponding low-SNR data we randomly reduced the detections by a factor of 20, using a binomial distribution for each pixel with $p=0.05$.
We use every 4th image as test set and keep the rest as training data for the supervised baseline.
This amounts to a total of 33 images (24 for training and 9 for testing) of size 280$\times$280 pixels.

\noindent {\bf The \emph{MT-SM} dataset} was derived from single molecule localisation data published by  Jimenez~\etal~in~\cite{jimenez2020samples}.
It shows the arranged cells labeled for microtubules.
To create the dataset, we binned the detected emitter locations using a bin size of 28nm $\times$ 28nm to produce the high-SNR data. 
To produce the corresponding low-SNR data we randomly reduced the detections by a factor of 200, using a binomial distribution for each pixel with $p=0.005$.
We use every 4th image as test set and keep the rest as training data for the supervised baseline.
This gives a total of 120 images (90 for training and 30 for testing) of size 640$\times$640 pixels.

% ...................................................
\subsection{Denoising Performance}
% ...................................................
To evaluate the denoising performance of our method we train one network for our method and one for each baseline (N2V2, HDN, and HDN256). Since these methods do not require clean data, we can train them on the full low-SNR data, including the section used for testing.
The supervised baseline, which requires clean training data, is trained only on the designated training section of the data.
Quantitative and qualitative results can be found in table~\ref{tab:psnr} and Figure~\ref{fig:resultsm}.
\tabpsnr

We find that our method is on-par or outperforms the baselines and even the supervised approach.
We believe that the reason for this might be that, depending on the data split, supervised methods might suffer from a mismatch between training and test distributions, which might be especially the case for the \emph{Conv-PC} dataset where test and training data consist of different FOVs showing slightly different patterns.

% ...................................................
\subsection{Diversity Denoising}
% ...................................................
In Figure~\ref{fig:divdenoising}, we qualitatively evaluate the performance of our method for diversity denoising, that is, its ability to sample diverse possible clean images from single noisy input.
To show the full range of possible results, we trained our method on the high-SNR data of the \emph{Conv-PC} dataset.

We generate six different shot noise corrupted versions of an image at different noise levels/photon numbers and use them as input for the sampling procedure described in Figure~\ref{fig:explainer}, to generate three possible clean versions for each noisy input image.
Noisy images with low photon counts can be explained by a broad range of possible clean images and yield highly diverse results.
Increasing the photon count of the input image, we find that the differences in the sampled clean images become more subtle until only local structures differ.
\figdivDenoising

% ...................................................
\subsection{Image Generation Performance}
% ...................................................
\figsamples
Finally, we evaluate our method for the use as a generative image model.
We are especially interested in the setting where only low-SNR data is available for training and want to investigate how the distribution of the generated images will compare to the clean high-SNR data.
We train our model as well as the HDN and HDN256 baselines on the low-SNR data for each dataset.
We generate 10000 images of 256$\times$256 pixels using our method (Figure~\ref{fig:explainer}) and HDN256. We then compute the FID~\cite{heusel2017gans} score against 10000 random crops of the augmented high SNR-data.
We compute the scores using the \emph{clean FID}~\cite{parmar2021cleanfid}.
For a fair comparison against the HDN baseline, trained on 64$\times$64 pixel patches, we compute the FID using 64$\times$64 pixel patches against 64$\times$64 crops of the high-SNR data.
To compare against our method, we use random 64$\times$64 crops from the 256$\times$256 pixel patches generated by our method.
Quantitative results can be found in Table~\ref{tab:fid}. 
Qualitative results for 256$\times$256 patches are shown in Figure~\ref{fig:samples}.
We find that our method visually outperforms HDN and HDN256 and consistently achieves lower FID scores for 256$\times$256 patches. 
For the smaller 64$\times$64 patches FID results are less clear.
We believe, that this is due to the fact, that larger structures are not captured at this patch size, and that our high-SNR data contains residual noise, which seems to be better represented by HDN.
\tabfid

% ---------------------------------------------------
\section{Discussion and Conclusion}
% ---------------------------------------------------
We have introduced a new perspective on shot noise-affected imaging and showed that it can be utilised for self-supervised MMSE denoising, obtaining diverse denoising solutions, and constructing generative models that can be trained with noisy data.
We believe that this perspective might open the door to new applications in areas of microscopy where only shot noise-affected data is available.
We also believe that our method can be extended to be used in a conditional setting for image-to-image translation, such as the prediction of fluorescence channels from bright-field images -- a topic that has received much attention in the recent years~\cite{lee2018deephcs,tonks2023evaluation}. 
While our method is currently limited to data purely affected by shot noise, we hope that future work can extend the approach to be applicable in a more general setting.
Finally, we applied GAP to two natural image datasets (see Supplementary material) with encouraging generative visual results. 
We believe GAP might be applicable as a generative model beyond microscopy.

% ---------------------------------------------------
\subsection*{Acknowledgements}
% ---------------------------------------------------
We would like to thank Jeremy Pike for pointing us the single molecule localisation data and for the helpful discussions we had.
The computations described in this paper were performed using the University of Birmingham's BlueBEAR HPC service, which provides a High Performance Computing service to the University's research community. See \emph{http://www.birmingham.ac.uk/bear} for more details.
We used CaStLeS~\cite{thompson2019castles} and \emph{Baskerville} resources.

{\small
\bibliographystyle{unsrt}
\bibliography{refs}
}

\ifarXiv
    \foreach \x in {1,...,\numbersupplementpages}
    {
        \clearpage
        \includepdf[pages={\x}]{\supplementfilename}
    }
\fi

\end{document}

%%%%%%%%% TITLE
\title{Image Denoising and the Generative Accumulation of Photons \\ 
(Supplementary~Material)}

\author[1]{Alexander~Krull}
\author[2]{Hector~Basevi}
\author[1]{Benjamin~Salmon}
\author[3]{Andre~Zeug}
\author[3]{Franziska~Müller}
\author[1]{Samuel~Tonks}
\author[1]{Leela~Muppala}
\author[1]{Ale\v{s}~Leonardis}

\affil[1]{Computer Science, University of Birmingham, UK}
\affil[2]{Metabolism and Systems Research, University of Birmingham, UK}
\affil[3]{Medizinische Hochschule Hannover, Germany}

\maketitle
\thispagestyle{empty}

% %%%%%%%%% ABSTRACT
% \begin{abstract}

% \end{abstract}
%%%%%%%%% BODY TEXT
\vspace{-3mm}

% ---------------------------------------------------
\section{Code and Data Availability}
% ---------------------------------------------------
The code and datasets will be made publicly available here: 

\noindent \emph{https://github.com/krulllab/GAP}.

% ---------------------------------------------------
\section{Comparing DDPM versus our Generative Model}
% ---------------------------------------------------
In Figure~\ref{fig:ddpm}, we qualitatively compare our results against a denoising diffusion model (DDPM)~\cite{ho2020denoising}.
While DDPM models achieve impressive sample quality, they are, unlike our method not able to learn the generation of denoised images when trained on noisy data.
\figddpm

% ---------------------------------------------------
\section{Additional Details on the Training Procedure}
% ---------------------------------------------------
During training, we must ensure that our CNN is able to produce high quality predictions for a range of photon counts.
To achieve this, we randomly pick a value for $p$ for each training patch (see section~3.5 in the main paper).
To achieve a good coverage over different noise levels/photon counts we use a concept we call pseudo-PSNR, explained in section~\ref{sec:psudopsnr}.
We sample uniformly from a range of pseudo-PSNR numbers and then compute the corresponding value for $p$ accordingly.
The process is described in section~\ref{sec:binomsampling} 

% ...................................................
\subsection{The Pseudo-PSNR value}
\label{sec:psudopsnr}
% ...................................................
% We use pseudo-PSNR as a concept to achieve a good cover of different noise levels, not as part of our evaluation.
When imaging a static sample the resulting PSNR number of the recorded image depends on the amount of light that was allowed to hit the detector.
We can expect a longer exposure time or stronger light intensity to produce a cleaner image with a higher PSNR value than an image recorded with a shorter exposure or reduced light intensity, in which the detector was allowed to collect fewer photons.
We define the intensity of an image as the average number of photons per pixel
\begin{equation}
    \gamma = \frac{|\obs|}{n}
    .
\end{equation}

To compute the PSNR value of a noisy image $\obs$ one generally requires the correctly scaled version of the normalised clean ground truth image $\sig$.
We can compute the correctly scaled signal as $\bar{\sig} = \gamma n \sig $, which is then directly comparable to the noisy image $\obs$.
The equation used for this is
\begin{equation}
    \mbox{PSNR}(\obs,\bar{\sig}) = 10 \log_{10} \frac{\bar{\sig}_{\mbox{\tiny max}}^2}{\mbox{MSE}},
    \label{eq:psnr}
\end{equation}
where $\bar{\sig}_{\mbox{\tiny max}}$ is the maximum value of the absolute signal $\bar{\sig}$, and ${\mbox{MSE}}$ is the mean squared error between $\obs$ and $\bar{\sig}$.

The idea of pseudo-PSNR is to directly compute the PSNR value we might expect for a shot noise corrupted image of a certain intensity, without requiring us to compare a noisy and clean image.
We define the pseudo-PSNR value for intensity $\gamma$ as the PSNR value we would expect for the shot noise corrupted version of a flat signal $\sig$, with all pixel values being $\sigpix_i = \frac{1}{n}$.
Based on the shape of the Poisson distribution, we should expect for such an image ${\mbox{MSE}}= \gamma$ and $\bar{\sig}_{\mbox{\tiny max}}= \gamma$.
Based on Eq.~\ref{eq:psnr}, we calculate the pseudo PSNR as
\begin{equation}
\begin{split}
    \mbox{PSNR}_{\mbox{\tiny PS}}(\gamma) & = 10 \log_{10} \frac{\bar{\sig}_{\mbox{\tiny max}}^2}{\mbox{MSE}} \\
    & = 10 \log_{10} \frac{\gamma^2}{\gamma} \\
    & = 10 \log_{10} \gamma
\end{split}
\label{eq:pseudopsnr}
\end{equation}
We can invert Eq.~\ref{eq:pseudopsnr} to compute the corresponding intensity $\gamma$ for a given pseudo PSNR value as
\begin{equation}
    \gamma = 10^\frac{{\mbox{PSNR}}_{\mbox{\tiny PS}} }{10}
    .
    \label{eq:psnrinvert}
\end{equation}

% ...................................................
\subsection{Training Pair Sampling}
\label{sec:binomsampling}
% ...................................................
Before we can split out training patches into input and target using a binomial distribution (see section~3.5 in the main paper), we have to determine the success probability parameter $p$ of the distribution.
To achieve this, for each training patch, we first sample from a uniform distribution over pseudo PSNR values between a predefined minimum and maximum.

The goal is to set $p$, so that the average photon number (intensity) of the resulting input image corresponds to the drawn pseudo PSNR value.
We compute the corresponding intensity $\gamma$ using Eq.~\ref{eq:psnrinvert} from the randomly determined pseudo PSNR value.
Then, we compute the corresponding success probability for the binomial distribution as
\begin{equation}
    p = \frac{\gamma}{|\obs|/n},
\end{equation}
such that the input image photon count after the split will correspond to the drawn pseudo PSNR.
The result is then clipped to values below $0.99$ to guarantee that at least 1\% of photons is on average assigned to the target image.

We use the following intervals to sample the pseudo PSNR values: $[-40:-5]$ for NPC-SM, $[-40,-10]$ for MT-SM, $[-40,-10]$ for Conv-PC, and $[-40,20]$ for Neuro-PC.

% ---------------------------------------------------
\section{Details on Photon Sampling Procedure}
% ---------------------------------------------------
Here, we want to discuss the details of the photon sampling procedure.
Depending on whether we perform image generation or diverity denoising, we initialise the process with an empty image or a noisy image $\obs_0$.
We then apply our trained CNN to compute the probability distribution over the possible next photon positions
\begin{equation}
    \bar{\sig}_t = f(\obs_t;\pars).
\end{equation}
Because of our softmax output layer it is guaranteed that each pixel value  $\bar{\sigpix}_{t,i} \geq 0$ and that 
\begin{equation}
    \sum_i^n \bar{\sigpix}_{t,i} = 1
    .
\end{equation}
We then sample a set of new photons represented by the image $\obs_t^{\mbox{\scriptsize new}}$, where each pixel value $\obs_{t,i}^{\mbox{\scriptsize new}}$ holds the number of photons that will be added at location $i$.
Each pixel value $\obspix_{t,i}^{\mbox{\scriptsize new}}$ is drawn from a Poisson distribution with mean 
\begin{equation}
    \lambda_{t,i} = n \bar{\sigpix}_{t,i} \alpha_t,
\end{equation}
where $\alpha_t$ controls how many photons will on average be sampled in total in $\obs_{t}^{ \mbox{\scriptsize new}}$.
We set as
\begin{equation}
     \alpha_t = \max (\beta \sum_i^n \obspix_{t,i}, 1),
\end{equation}
where the parameter $\beta$ controls the rate at which the photon number increases on average.
In our experiments, we set $\beta= 10\%$.
The maximum operation ensures that the number of photons is increasing from the beginning even when starting with an empty image $\obs_{0}$.
Finally, we compute the next photon count image as
\begin{equation}
    \obs_{t+1} = \obs_t + \obs_{t}^{\mbox{\scriptsize new}}
\end{equation}
and repeat the process.

% ---------------------------------------------------
\section{Detailed Description of Datasets}
% ---------------------------------------------------
Here, we want to give additional information about the photon counting datasets we recorded.

\subsection{Conv-PC dataset}
Data was recorded from a Convallaria majalis rhizome section sample slide at a Leica TCS SP8 TPE DIVE with FALCON and the HC PL IRAPO 25x dipping objective. 
We used 850 nm excitation at 1\% laser power, HyD-RLD detector, emission range of 600 – 650 nm, pixel size 0.6 x 0.6 µm, an 8 MHz resonant scanner and 4x averaging.
We imaged 5 fields of view (FOVs), each containing 512x512x2048 voxels (xyt). Four FOVs were used as training data for supervised denoising, and the fifth FOV was used as test data.

We used a time binning of 4, resulting in datasets of 512x512x512 voxels, and named these low-SNR raw data 'trainingData.tif' and 'testData.tif' respectively.
Furthermore, we summed these 512 frames to produce the high SNR ('ground truth') version and named it 'trainingDataGT.tif' and 'testDataGT.tif'.
Each frame of the raw data contains a significant amount of image noise. All voxel values correspond to photon counts.

\subsection{Neuro-PC dataset}
Data was obtained from 11–14 week old male Mice (C57BL/6J background) stereotactically injected with AAV-hGFAP-5-HT4R-eGFP and AAV-hGFAP-tdTomato to the CA1 region of the hippocampus 3 weeks prior to experiment. 
Data was recorded from acute slices of the mouse hippocampus region at a Leica TCS SP8 TPE DIVE with FALCON, using following acquisition settings: HC PL IRAPO 25x dipping objective, excitation 920 nm at 15-30\%, HyD-RLD detector, emission 490 – 560 nm (eGFP), and 560 – 650 nm (tdTomato), voxel size ( 0.1 x 0.1 x 0.5 µm), 4x averaging, scan speed 600 Hz.
We are using only the tdTomato channel.

% ---------------------------------------------------
\section{Network Architecture}
% ---------------------------------------------------
We use a modified UNet~\cite{ronneberger2015u} architecture, with skip connections and residual blocks.
Each residual down-block and up-block consists of 3 3$\times$3 convolutions with RELU activation functions after the second convolutions and at the end of the block.
We use max-pooling for down-sampling and transposed convolutions for up-sampling.

Since we are training a single network to handle a range of different noise levels and photon counts in its input, normalizing the input is not trivial.
To avoid normalization, we use a sinusoidal frequency encoding~\cite{tancik2020fourier} applied to each pixel value at the input of our network.
We use 10 different sinosoids with frequencies at different powers of 10.

% ---------------------------------------------------
\section{Hyper Parameters}
% ---------------------------------------------------
Since our datasets have differing sizes, we define one training epoch as 500 training steps.
We use the first 90\% of images in each dataset as training data and the last 10\% as validation set.
We use the ADAM optimizer~\cite{kingma2015adam}.
We use an initial learning rate of 1e$-4$ and reduce the learning rate using the pytorch \emph{ReduceLROnPlateau} scheduler with a patience of 10 by a factor of $2$.

% ---------------------------------------------------
\section{Details on the Loss Function}
% ---------------------------------------------------
Here we show that the loss function from Eq.~13 in the main paper, which uses target images $\obs^k_{{\tiny \mbox{tar}}}$ with multiple photons is equivalent to using single photons represented by one-hot-encoding images. 
We can write the loss as
\begin{equation}
\begin{split}
     L(\pars) &= -
    \sum_{k=1}^m 
    \sum_{i=1}^n
    \ln f_i(\obs^k_{\tiny \mbox{inp}};\pars)
    \obspix^k_{{\tiny \mbox{tar}},i}   
    \\
    &= -
    \sum_{k=1}^m 
    \sum_{i=1}^n
    \ln f_i(\obs^k_{\tiny \mbox{inp}};\pars)
    \sum_{t=1}^T  \obspix^{k,t}_{{\tiny \mbox{tar}},i}  ~~~~,
\end{split}
    \label{eq:loss}
\end{equation}
where $\sum_{t=1}^T  \obspix^{k,t}_{{\tiny \mbox{tar}},i}$ is the one-hot-encoding photon image for photon $t$ from $\obs^k_{{\tiny \mbox{tar}}}$.
Note that the order of photons does not matter here.
We can then continue to write
\begin{equation}
\begin{split}
     L(\pars) &= -
    \sum_{k=1}^m 
    \sum_{i=1}^n
    \ln f_i(\obs^k_{\tiny \mbox{inp}};\pars)
    \sum_{t=1}^T  \obspix^{k,t}_{{\tiny \mbox{tar}},i}   
    \\
     &= -
    \sum_{k=1}^m 
    \sum_{t=1}^T 
    \sum_{i=1}^n
    \ln f_i(\obs^k_{\tiny \mbox{inp}};\pars)
    \obspix^{k,t}_{{\tiny \mbox{tar}},i}  
    ~~~~.
\end{split}
\end{equation}
In this formulation it becomes clear that using target images with multiple photons is equivalent to replicating each input image $T$ times and using it together with each of the corresponding single-photon target images.
This corresponds to the same training data distribution as randomly sampling single photon targets.

% ---------------------------------------------------
\section{Additional qualitative results}
% ---------------------------------------------------
We show randomly selected samples for all datasets in Figure~\ref{fig:samplesfull}.
\figsamplesFull

% ---------------------------------------------------
\section{Natural image datasets}
% ---------------------------------------------------
To demonstrate the potential of our method, we show randomly selected outputs of our generative model when applied on natural image datasets in Figure~\ref{fig:nares}.
To account for the greater complexity of these datasets, we trained 8 expert networks, each specialised on a sub-range of pseudo PSNR values. 
Each network is scaled up to 7 levels (instead of 6) and starting with 32 feature (instead of 28) channels.
When generating images we switch between these expert networks as the image gains more and more photons.
Apart from this, the approach is the same as for the microscopy data.
\fignatresults

\clearpage
{\small
\bibliographystyle{unsrt}
\bibliography{refs}
}